\documentclass[preprint]{aastex}
\usepackage[utf8x]{inputenc}
\usepackage[OT1]{fontenc}
\usepackage[english]{babel}
\usepackage{hyperref}
\usepackage{subfig}
\usepackage{graphicx,latexsym,layout,dsfont,amsfonts,amsmath,amssymb,fontenc,xcolor,wrapfig,multirow,amsfonts,bm,textcomp,epstopdf}
\newcommand{\pam}{\textsf{PAMELA}}

\begin{document}


\title{Observations  of  the  December  13 and 14, 2006,  Solar  Particle  Events in the 80 MeV/n - 3 GeV/n range from  space with  \pam\  detector.}
\author{O. Adriani,\altaffilmark{1, 2} G. C. Barbarino,\altaffilmark{3, 4}
G. A. Bazilevskaya,\altaffilmark{5} R. Bellotti,\altaffilmark{6,
7} M. Boezio,\altaffilmark{8} \\
E.A. Bogomolov,\altaffilmark{9} L. Bonechi,\altaffilmark{1, 2} M.
Bongi,\altaffilmark{2} V. Bonvicini,\altaffilmark{8} S.
Borisov,\altaffilmark{10, 11, 12}
S.Bottai,\altaffilmark{2} \\
A. Bruno,\altaffilmark{6, 7} F.
Cafagna,\altaffilmark{7} D. Campana,\altaffilmark{4} R.
Carbone,\altaffilmark{4, 11} P. Carlson,\altaffilmark{13}
M.Casolino,\altaffilmark{10,18} \\
G. Castellini,\altaffilmark{14} L.
Consiglio,\altaffilmark{4} M. P. De Pascale,\altaffilmark{10, 11}
C. De Santis,\altaffilmark{10, 11}
N. De Simone,\altaffilmark{10, 11} \\
V. Di Felice,\altaffilmark{10,11} A. M. Galper,\altaffilmark{12} L.
Grishantseva,\altaffilmark{12}
W. Gillard,\altaffilmark{13} G. Jerse,\altaffilmark{8, 15} A. V. Karelin,\altaffilmark{12} \\
S. V. Koldashov,\altaffilmark{12 }S. Y. Krutkov,\altaffilmark{9 }A. N.
Kvashnin,\altaffilmark{5} A. Leonov,\altaffilmark{12}, V. Malakhov, \altaffilmark{12} L. Marcelli,\altaffilmark{10}\\
A.G. Mayorov,\altaffilmark{12} W.Menn,\altaffilmark{16} V. V. Mikhailov,\altaffilmark{12} E. Mocchiutti,\altaffilmark{8}
A. Monaco,\altaffilmark{7} N. Mori,\altaffilmark{1,2} \\
N. Nikonov,\altaffilmark{9, 10, 11} G. Osteria,\altaffilmark{4} F. Palma,\altaffilmark{10,11} P.
Papini,\altaffilmark{2} M. Pearce,\altaffilmark{13} P. Picozza,\altaffilmark{10,11} \\
C. Pizzolotto,\altaffilmark{8} M. Ricci,\altaffilmark{17} S. B. Ricciarini,\altaffilmark{2} R. Sarkar,\altaffilmark{8} L.
Rossetto,\altaffilmark{13} M. Simon,\altaffilmark{16}  \\
R.Sparvoli,\altaffilmark{10, 11 \ }P. Spillantini,\altaffilmark{1,2}
Y. I. Stozhkov,\altaffilmark{5} A. Vacchi,\altaffilmark{8} E.
Vannuccini,\altaffilmark{2} G. Vasilyev,\altaffilmark{9} \\
S. A. Voronov,\altaffilmark{12} J. Wu,\altaffilmark{13}
Y. T. Yurkin,\altaffilmark{12} G. Zampa,\altaffilmark{8} N.
Zampa,\altaffilmark{8} and V. G. Zverev\altaffilmark{12}}

\altaffiltext{1}{University of Florence, Department of Physics,
I-50019 Sesto Fiorentino, Florence, Italy}

\altaffiltext{2}{INFN, Sezione di Florence, I-50019 Sesto Fiorentino,
Florence, Italy}

\altaffiltext{3}{University of Naples "Federico
II", Department of Physics, I-80126 Naples, Italy}

\altaffiltext{4}{INFN, Sezione di Naples, I-80126 Naples, Italy}

\altaffiltext{5}{Lebedev Physical Institute, Leninsky Prospekt 53,
RU-119991 Moscow, Russia}

\altaffiltext{6}{University of Bari, Department of Physics, I-70126
Bari, Italy}

\altaffiltext{7}{INFN, Sezione di Bari, I-70126 Bari, Italy}

\altaffiltext{8}{INFN, Sezione di Trieste, I-34149 Trieste, Italy}

\altaffiltext{9}{Ioffe Physical Technical Institute, RU-194021 St.
Petersburg, Russia}

\altaffiltext{10}{INFN, Sezione di Rome Tor Vergata, I-00133 Rome, Italy}

\altaffiltext{11}{University of Rome Tor
Vergata, Department of Physics, I-00133 Rome, Italy}

\altaffiltext{12}{National Research Nuclear University "MEPhI", RU-115409
Moscow, Russia}

\altaffiltext{13}{KTH, Department of Physics, and the Oskar Klein
Centre for Cosmoparticle Physics,
AlbaNova University Centre, SE-10691 Stockholm, Sweden}

\altaffiltext{14}{IFAC, I-50019 Sesto Fiorentino, Florence, Italy}

\altaffiltext{15}{University of Trieste, Department of Physics,
I-34147 Trieste, Italy}

\altaffiltext{16}{Universitat Siegen, Department of
Physics, D-57068 Siegen, Germany}

\altaffiltext{17}{INFN, Laboratori Nazionali di
Frascati, Via Enrico Fermi 40, I-00044 Frascati, Italy}
\altaffiltext{18}{RIKEN, Advanced Science Institute, Wako-shi, Japan}

\begin{abstract}
We present the space spectrometer \pam\ observations of
proton and helium fluxes during the December 13 and 14, 2006 solar
particle events. This is the first direct measurement of the solar
energetic particles in space with a single instrument in the energy
range from $\sim$ 80 MeV/n up to $\sim$ 3 GeV/n. In the event of December
13 measured energy spectra of solar protons and helium were compared with results
obtained by neutron monitors and other detectors. Our measurements show a spectral behaviour different from
those derived from the neutron monitor network. No satisfactory analytical fitting was
found for the energy spectra. During the first hours of the
December 13 event solar energetic particles spectra were close to the exponential form
demonstrating rather significant temporal evolution. Solar He with energy up to ~1 GeV/n was recorded on December 13.  In the event of December 14 energy of solar protons reached ~600 MeV whereas maximum energy of He was below 100 MeV/n. The spectra were slightly bended in the lower energy range and preserved their form during the second event. Difference in the particle flux appearance and temporal evolution in these two events may argue for a special conditions leading to acceleration of solar particles up to relativistic energies.
\end{abstract}

\keywords{ space spectrometer, solar
particle emission, energy spectrum}

\section{Introduction}

The \pam\ spectrometer \citep{picozza_pamela_2007} is a space instrument designed for the study
of primary charged particles and antiparticles in a wide energy
interval, mainly from tens of MeV to ${\approx}$ 1.2 TeV for protons. It was launched in an elliptical orbit at
an altitude between 350 and 610 km and inclination of 70 degree in June
2006. The main scientific goal of \pam\ is the measurement of the particle and antiparticle component in galactic cosmic rays \citep{adriani_anomalous_2009,
adriani_new_2009}, the study of galactic cosmic ray modulation by solar activity, and
solar energetic particles \citep{casolino_cosmic-ray_2006,de_simone_study_2009}. This paper reports the \pam\ measurements of the solar proton
and helium fluxes in the energy range from below 100 MeV/n to several
GeV/n during the December 13 and 14, 2006 solar particle events.

The problem concerning mechanism and site of solar energetic
particles (SEP) acceleration remains an open question. Certainly, SEP may be produced
 after powerful explosive events on the Sun,
accompanied by solar flares, Coronal Mass Ejections (CME), bursts of
solar X /gamma-rays and radio emission \citep{reames_particle_1999}. It
is clear that not a single mechanism is involved into the SEP generation.
Stochastic acceleration, shock acceleration, and acceleration by the
DC electric fields in the process of magnetic reconnection are the
main candidates. Acceleration of SEP may take place in the flare
region, solar corona and even in the interplanetary space. It should be
kept in mind that SEP themselves, while propagating, create conditions for the energy redistribution
\citep{tylka_new_2001,lee_coupled_2005}. Therefore, the energy spectrum of
SEP provides valuable information for study of solar and interplanetary plasma
processes. Most energetic particles are usually (not always) accelerated
during short time close to the energy explosion, thus the first
particles arriving at the observer site keep more information
about the primary acceleration process. Less energetic SEP may be
accelerated during hours and even days. Conditions for particle
acceleration in the corona and interplanetary space are determined by temporal and spatial evolution of a shock \citep[e.g.][]{ellison_shock_1985}.

Since the energy range of SEP extends over more than 5 orders of
magnitude, and the SEP fluxes occupy more than 8 orders of
magnitude, the SEP energy spectrum has never been measured by a
single instrument. The bulk of the SEP observations has been
performed on spacecraft and covers the particle energy below
several hundreds MeV/n, whereas the relativistic solar particles
have been observed by ground-based installations and remain less
elucidated. This kind of observations and associated modeling can
also be used to issue radiation alerts to be used by manned and
unmanned missions. A multi-detector study of the solar particle
event occurring in 2005 and the corresponding  GLE69 was performed
on-ground using NM and SREM  radiation environment units in space
\cite{astra-7-1-2011}

It is not yet clear if relativistic solar particles require a
special scenario for their generation. Some authors believe that
relativistic solar protons have two components - the fast and the
delayed ones. The fast component is initially accelerated in the
flare region
\citep[e.g.][]{vashenyuk_relativistic_2006,vashenyuk_characteristics_2008,vashenyuk_relativistic_2008,mccracken_two_2008},
probably via magnetic field reconnection, whereas another
mechanism, probably shock or stochastic acceleration, generates
the second particle component.
\cite{bieber_spaceship_2004,bieber_relativistic_2005} argue
against two components and in favour of the CME-driven shock as a
main accelerator of the whole SEP population. A wide variety in
concomitant acceleration conditions supports the point that
various mechanisms play the main role in different SEP events
\citep{bombardieri_improved_2008,bazilevskaya_early_2009}.
Detailed studies of more solar particle spectra in wide energy
range are needed to resolve this problem.

Since ground-based instruments can only detect secondary cosmic rays, knowledge of response to the primary particles is necessary in order to find SEP fluxes in space.
Some methods were developed \citep{shea_possible_1982,cramp_october_1997,bieber_spaceship_2004,plainaki_modeling_2007,vashenyuk_relativistic_2006} to derive energy spectra and angular anisotropy of solar particles using the enhancements in the count rates of neutron monitors. Nevertheless, the particle fluxes derived from the neutron monitor data are model dependent.

Before the \pam\ launch, direct measurements of the relativistic solar particles
were not fulfilled. The aim of this paper is to present the results of
the first direct measurements by a single instrument of the solar
protons in the energy range from 80 MeV to several GeV and the helium
nuclei from 75 MeV/n up to $\sim$ 700 MeV/n during the SEP events of
December 13 and 14, 2006. The absolute intensities and energy spectra
are compared with the results of direct (GOES, ACE) and indirect (neutron monitors, IceTop) measurements.

\section{Instrument Description}

\begin{figure}[h]
  \begin{center}
  \includegraphics[width=\textwidth]{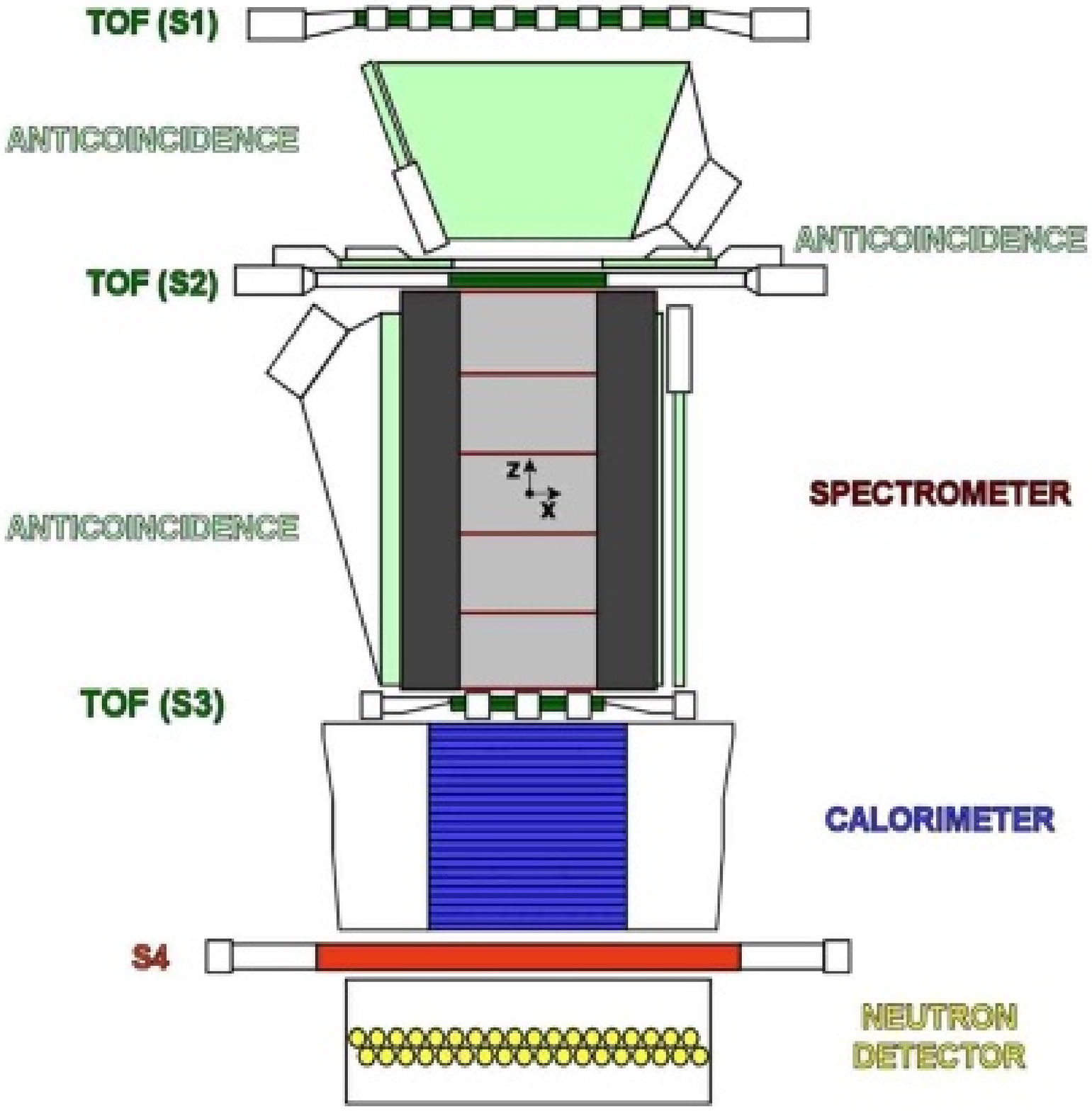}
  \end{center}
  \caption{Scheme of the \pam\ instrument. S1, S2, S3 - TOF scintillator planes (each plane consists of 2 layers), CARD, CAT, CAS - anticoincidence system, Spectrometer - tracker surrounded with a permanent magnet, Silicon-Tungsten calorimeter, S4 - Shower tail catcher, neutron detector.  Particles enter the detector from the top, cross the scintillators and are bent in the magnetic spectrometer before interacting with the calorimeter.}
  \label{fig_pam}
\end{figure}

The \pam\ spectrometer (Figure \ref{fig_pam}) consists of a number of highly
redundant detectors capable of particles detection trough the
determination of their charge, mass, rigidity and
velocity over a very wide energy range. Total weight
of \pam\ is 470 kg; power consumption is 355 W.

A more detailed description of the device and the data handling can be
found in \citep{picozza_pamela_2007,casolino_pamela_2006,casolino_yoda++:_2006}. The core of the instrument is a
spectrometer, constituted by permanent magnet with a silicon microstrip tracker
providing momentum and charge (with sign) information
\citep{adriani_powerful_2003}. The permanent magnet is 43.66 cm high and
constructed of 5 modules with high residual magnetic induction (0.43 T)
providing an almost constant magnetic field value in the cavity. The
tracking system \citep{ricciarini_pamela_2007} is composed of 6 planes of
high-precision Si microstrip detectors, positioned between the 5
magnetic modules of the tower, with uniform vertical spacing of 8.9
cm. Each silicon plane performs measurements of energy release and track position with a
precision of about $3.0\; \mu m$ (X - bending view) and $11.5\; \mu m$ (Y
view).

A scintillator system \citep{barbarino_pamela_2003} provides trigger, charge
and time of flight (TOF) information. It is composed of three double
layers S1, S2, and S3, divided in various bars for a total of 48
scintillator paddles. It is also used to reject albedo particles which cross
the detector from bottom to top.

An anticoincidence system (CARD, CAT, and CAS) is used to reject spurious events in the off-line phase.

A silicon-tungsten calorimeter \citep{boezio_high_2002} is used to perform
hadron/lepton separation. It is composed of 44 silicon layers
interleaved by tungsten planes for a total of $16.3\;X_0$ (radiation lengths) and $0.6 \; \lambda_i$ (nuclear interaction lengths).

A shower tail catcher S4 and a neutron detector at the bottom of the
apparatus are used to increase lepton/hadron separation.

Around the detectors are housed the readout electronics, the interfaces
with the CPU and all primary and secondary power supplies. All systems
are redundant with the exception of
the CPU which is more tolerant to failures. The system is enclosed in a
pressurized container located on one side of the Resurs-DK1 satellite.

\section{Data Selection and Analysis Criteria}

\subsection{Geometrical Factor and Top of the Payload correction}
 The Geometrical factor ($G_f$) of the detector has been evaluated defining a
 fiducial area consisting of a frame of $0.15 ~cm$ from the walls of the $13.1\times 16.1~cm^2$ wide magnetic cavity. Only particles inside this fiducial area have been selected. This reduced volume ensures that all particles entering the magnetic cavity
 cross the scintillators and do not hit the magnet walls. The value of $G_f=19.93$ $
cm^2 sr$, constant within $1 \%$, has been estimated
 with a numerical calculation and cross-checked with a Monte Carlo simulation where all the physical interaction processes are
inactive, with the exclusion of particle bending in the magnetic
field. The two values of $G_f$ have been found in agreement within $0.1\%$.

Interactions losses, due to local interactions where taken into account as a scale factor added to the Geometrical Factor. Protons and helium nuclei may be lost due to scattering and/or hadronic interactions in the $2~mm$ Al thick pressurized container in which \pam\ is housed or in the top of the detector.
Correction factors amounts to $\simeq 6\%$ for protons and $\simeq 12\%$ for He due to the different cross section of the two species.

\subsection{Trigger system}

In low radiation regions, close to the geomagnetic equator and outside the South Atlantic Anomaly, where \pam\ crosses the trapped protons of the inner Van Allen belt, particles must cross at least one of the two layers of the three scintillator systems ($S1$, $S2$, $S3$) to provide a valid trigger. In low cutoff regions or in the Van Allen belts, where particle rate is higher, $S1$ signal is not required to avoid random triggers due to the high number of particles.
Given the low energy of solar particles compared to the galactic nuclei, in this analysis we have used particles selected in high latitude regions and thus taken with only ($S2$ and $S3$) configuration.

\subsection{Live Time}

 The live time $t_{live}$
 of the apparatus is evaluated using the scintillator and trigger system.
 The counters for the live and dead time ($t_{live}$, $t_{dead}$) are cross-checked with the on-board time of the CPU, measuring the
 acquisition time ($t_{acq}=t_{live}+t_{dead}$), to remove
 possible systematic errors due the counting method. The error associated with clock resolution is negligible compared to other sources of systematic errors.

\subsection{Event Selection}

\subsubsection{Time of flight system selection}

In this analysis we have selected events that do not produce secondary particles in the $S1$ and $S2$ scintillators and in the tracker, requiring a single fitted track within the spectrometer fiducial acceptance and a maximum of one paddle hit in $S1$ and $S2$ matching the extrapolated from the tracker trajectory.

Particles interacting in the satellite can produce showers with
different secondaries hitting the scintillator pads. These showers
may produce random coincidences in the scintillators: they are rejected
by means of anticoincidence and TOF cuts. We require the absence of hits in the anticoincidences around
the empty area between $S1$ and $S2$ and around the top magnetic
cavity (CARD and CAT respectively).
No constraints
on the anticoincidences around the magnet (CAS) and $S3$ have been put, since they are more often
hit by backscattered secondaries produced in the calorimeter. The
probability for such particles to hit CARD and CAT has
been estimated with experimental data and cross-checked with
Monte Carlo simulations. This efficiency has been included in the proton flux estimation.
No constraints on the anticoincidence have been imposed for helium.

The timing information of the TOF
scintillator paddles along the extrapolated trajectory is used to evaluate the $\beta$ of the particle.
Albedo particles crossing the detector from bottom to top are
discarded by requiring a positive $\beta $.

\subsubsection{Proton and Helium identification}

 Since average energy
loss of a charged particle through matter follows Bethe Bloch formula, $dE/dx
\varpropto Z^2/\beta^2$ (neglecting logarithmic terms), the
measurement of the average energy released in the tracker planes
for a given event at a given rigidity can be used to identify different particles.
Proton and helium candidates have been selected requiring
 energy loss in the
tracker planes compatible with Z=1 and Z=2 nuclei.

\begin{figure}
  \includegraphics[width=\textwidth]{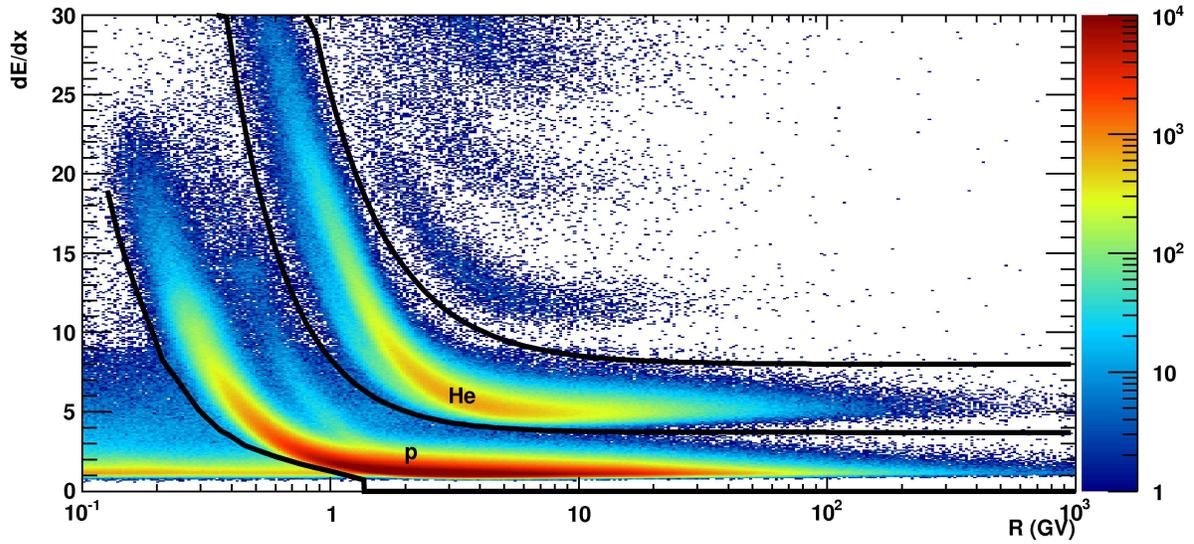}
  \caption{The energy loss (dE/dx) vs rigidity in the \pam\ tracker. The black lines show the rigidity dependent cuts used to select proton and Helium samples.}
\label{fig_dedx}
\end{figure}

Cuts in the energy loss (dE/dx) vs. rigidity remove positrons,
pions and particles with $Z\geq 2$ as shown in Figure \ref{fig_dedx}.
 The bands in the Figure due to protons and helium nuclei which have energy loss in the tracker $Z^2=4$
times protons, are identified. Events with small energy losses
below 1 GV are due to positrons, relativistic also at low rigidities
and the background of secondary pions. The contribution from these two particles are negligble respect to protons above 1 GV. The black
lines show the rigidity dependent cuts used to select the proton and helium
samples. From the same figure the deuterium contribution at low rigidities can also be identified, resulting in a band with
energy releases higher than protons. In this work we did not discriminated between isotopes.

Using the redundancy of energy loss measurements in tracker and TOF, residual contamination of protons in the helium sample has been found to be below 0.5$\%$ in the energy range used for this work.

\subsubsection{Geomagnetic Selection}

The high
inclination ($70^{\circ}$) orbit of the Resurs-DK1 satellite allows particles of different origin and nature to be studied. To separate the primary (solar and
galactic) component from the reentrant albedo component (particles produced
in cosmic ray interactions with the atmosphere below the cutoff
and propagating along Earth's magnetic field line), we evaluated the local geomagnetic cutoff $G$ in
the St\"{o}rmer approximation~\citep{shea_estimating_1987}. The value of $G=14.9/L^2$ - valid for vertically incident particles - is
estimated calculating the McIlwain $L$ shell with IGRF magnetic field model along the orbit~\citep{IGF2011}. Particles were selected requiring $R>1.3
\: G$ to remove any effect due to directionality in the detector and Earth's penumbral regions.

\subsubsection{Tracker selection}


Particle rigidity is obtained fitting it's track in the
spectrometer. In this analysis we selected events with a single track fully contained inside the fiducial acceptance.

For each particle, the tracking system provides up to 12 position measurements
(6 in the bending view), which are interpolated with a trajectory evaluated by integrating the
equations of motion in the magnetic field.
At low rigidies (below $\simeq  1 \:GV $) we have corrected for the energy loss in the detector.

The Maximum Detectable Rigidity (MDR) for a given detector is defined as the rigidity for which the relative error on the rigidity $\Delta R/R=100\%$ and varies between 200 GV and 1.5 TV, depending on delta-ray production and event's topology.
In this analysis, therefore, the particle rigidity is well below the MDR and thus is no MDR requirements have been applied.

\subsubsection{Selection efficiencies and residual contamination}

The tracker efficiency has been measured selecting a sample of
events which leave straight tracks in the calorimeter and do not
interact hadronically. These tracks were propagated back through \pam\ acceptance and tracker efficiency has been evaluated.

\begin{figure}
  \includegraphics[width=\textwidth]{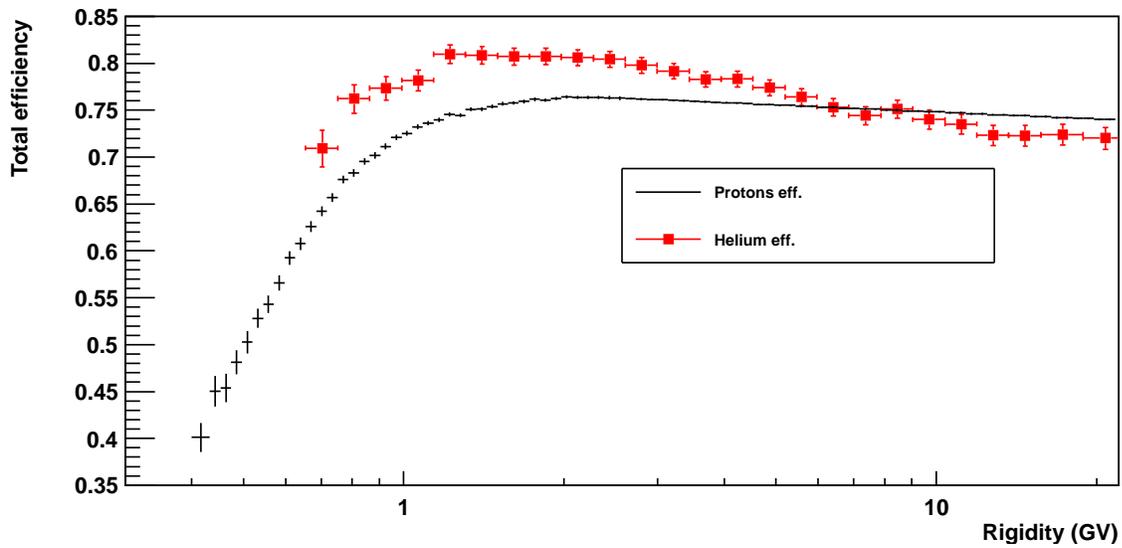}
  \caption{Selection efficiencies for proton and helium particles after all cuts are applied.}
\label{fig_eff}
\end{figure}

The total selection efficiencies for both species has been obtained as a product of Tracker, TOF (anticoincidence) efficiencies. The resulting efficiencies for proton and helium nuclei after all above mentioned cuts are shown in Figure \ref{fig_eff}. For a detail discussion see  \cite{adriani_pamela_2011}.

Although we put strong requirements on Tracker, TOF and anticoincidence systems, nevertheless there is residual contamination of secondary particles produced on the top of the payload entering the \pam\ acceptance window and passing selection cuts. The maximum contribution to the background for protons comes from secondary single-charge particles (positrons and pions). In case of helium, rejection power of our selection cuts is enough to make residual contamination negligible. In order to estimate this contamination we carried out a $2 \pi$ Monte Carlo simulation of the protons and helium nuclei hitting the \pam\ pressurized container. Two different hadronic interactions packages, based on Fluka \citep{battistoni_fluka_2007, ferrari_fluka:_2005} and Geant 4 \citep{allison_geant4_2006, agostinelli_g4--simulation_2003}, have been employed to simulate these interactions. This background decreases with increasing energy and amounts to less than few percents at 1 GV. Flux attenuation was estimated with an energy dependent simulation and it is constant above several GV.

\section{December 13 and 14, 2006 Solar Particle Events}

\subsection{General description}

The most significant SEP fluxes detected by \pam\,
were started on December 6, 13 and 14, 2006 and originated from the active
region NOAA 10930. December 6 event originated at the East limb
resulting in a gradual proton event reaching the Earth on December 7
and lasting until the events of December 13 and 14. Due to a
scheduled maintenance procedure, no data were collected by PAMELA tracker
during December 6 event, requiring a different analysis approach that will be discussed in a forthcoming paper. On December
13 2006, at 0214 UT an X3.4/4B solar flare occurred in the same active
region NOAA 10930 (S06W23) \citep{NOAA2011}. The intensity of
the event was quite unusual for a solar minimum condition.
This event produced also a full-halo Coronal Mass Ejection (CME) with the sky
plane projected speed of 1774 km/s \citep{adata2011}. The forward shock of the CME
reached Earth at 1438 UT on December 14, causing a Forbush decrease of
galactic cosmic ray intensities which lasted for several days.

The flare X1.5 (S06W46) at 2107 UT on December 14 gave
start to a new growth of particle intensity as recorded by \pam\ and
other satellites. The maximum energy of protons was below 1
GeV, therefore no ground level enhancement (GLE) was recorded.
The corresponding CME had a velocity of 1042 km/s.

\begin{figure}
  \includegraphics[width=1.15\textwidth]{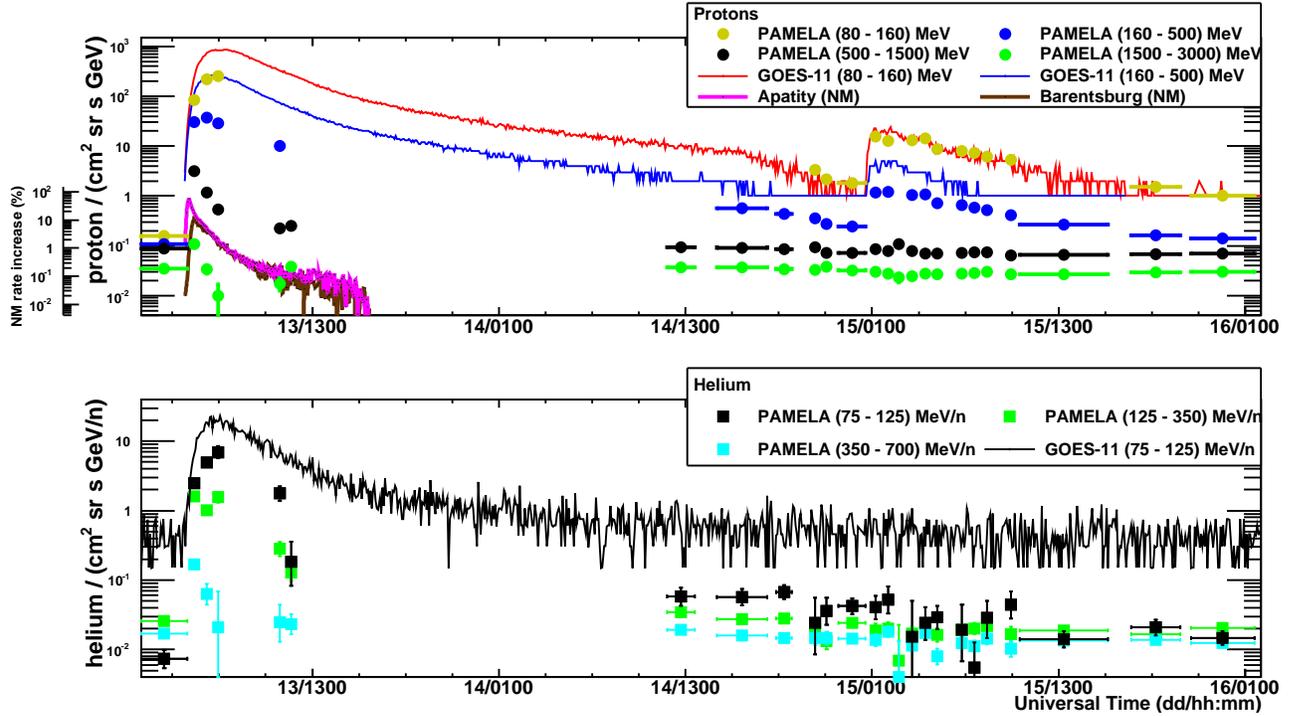}
  \caption{Intensity of solar protons (top panel) and helium (bottom panel)
of various energies as a function of time. Top panel: Circles are the PAMELA protons  in the ranges of 80-160 MeV (yellow), 160-500 MeV (blue), 500-1500 MeV (black), and 1.5-3.0 GeV. (green). Lines are the GOES 11 protons for  80-160 MeV(red) and 160-500 MeV (blue) ranges. The enhancement in the neutron monitor count rates in percents respect to the background before the event is given for Apatity (Rc=0.65 GV, violet line) and for  Barentsburg (Rc=0.1 GV, brown line). Bottom panel: Squares are the PAMELA helium in the ranges  of 75-125 MeV/n (black), 125-350 MeV/n (green), and 350-700 MeV/n (cyan). Black line is 75-125 MeV/n He from GOES 11.}
 \label{fig_time_profile}
\end{figure}

Figure \ref{fig_time_profile}
shows the solar particles intensity time profiles with
various energies as observed by the \pam\ spectrometer, GOES-11 and neutron monitors.
Ground-based neutron monitors started recording of SEP
at 0248 UT, while the near-Earth satellites responded later to SEP of lower
energies. First solar protons arrived at the Earth's
orbit with anisotropic pitch-angular distribution which resulted in
different intensity-time profiles at neutron monitors situated at
location with the same geomagnetic cutoff $R_c$ but
different asymptotic directions of incoming particles.
A highly collimated bunch of relativistic solar particles was observed by the
muon hodoscope URAGAN at 0256-0304 UT \citep{timashkov_ground-level_2008}.
Anisotropy vanished by 0430 UT \citep{vashenyuk_characteristics_2008}.

The whole energy range of SEP detectable by the \pam\
spectrometer can only be taken in the high latitude regions,
where the geomagnetic cutoff is lower than the rigidity of solar particles.
Measurements were performed during 5 passages of \pam\ through the polar regions on December
13. Data were missed from 1000 UT of December 13 till 0914 UT of
December 14 because of onboard system reset of the satellite.

\begin{figure}
  \begin{center}
  \includegraphics[width=1.1\textwidth]{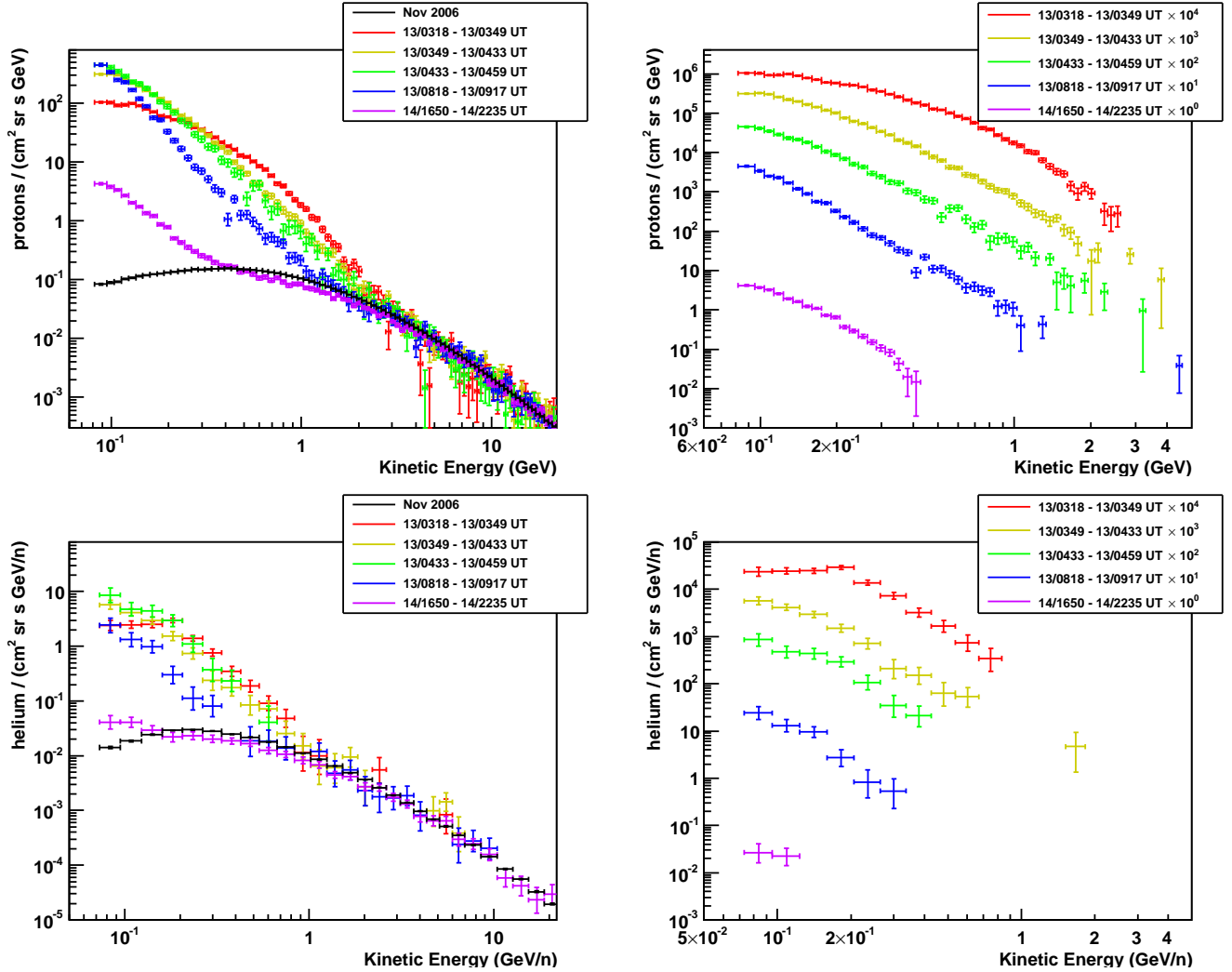}
  \end{center}
  \caption{Proton (top panels) and helium (bottom panels) spectra as measured by \pam\ for the SEP event of December 13, 2006. Left: Absolute fluxes, Right: After subtraction of galactic background. Fluxes are scaled to improve separation of curves. In the time slices of Dec 14, 1650-2235 UT it is possible to see the Forbush decrease of galactic particles after the arrival of the CME.}
\label{fig_13}
\end{figure}

\begin{figure}
  \begin{center}
  \includegraphics[width=1.1\textwidth]{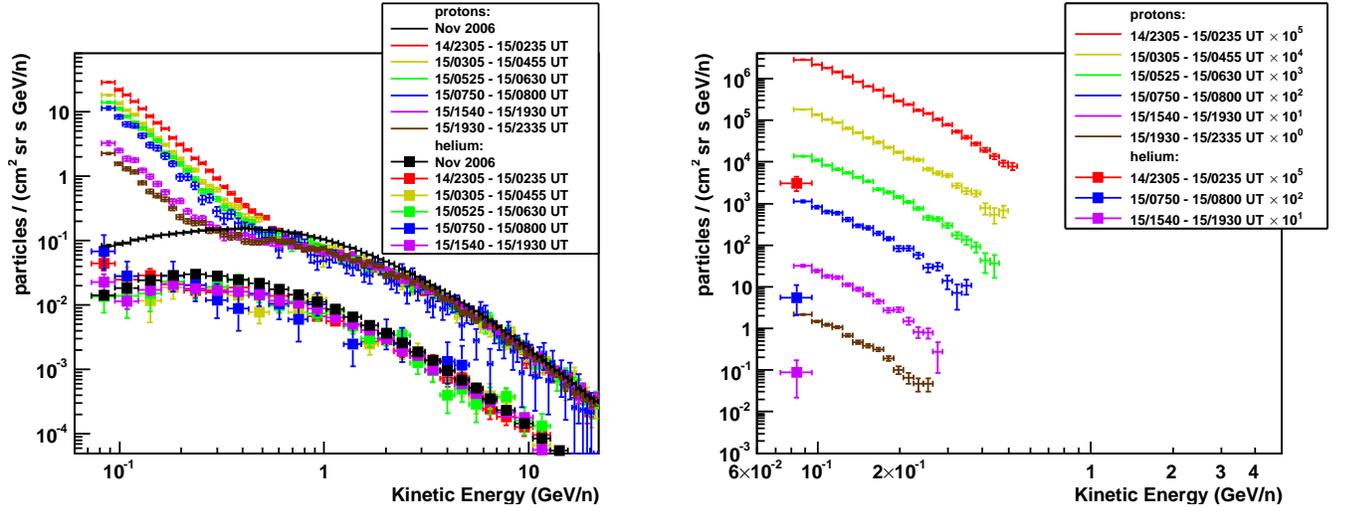}
  \end{center}
  \caption{Proton and helium spectra as measured by \pam\ for the SEP event on 14 December 2006. Left: Absolute fluxes, Right: After subtraction of galactic background. Fluxes are scaled to improve separation of curves.}
\label{fig_14}
\end{figure}

The results of \pam\ measurement during the December 13 and 14, 2006
SEP events are shown in Figure \ref{fig_13} and \ref{fig_14}
respectively. The quiet time galactic spectra were measured in
November 2006. The first (southern) polar passage of \pam\ started at
0318 UT and showed strongly flattening of p and He spectra at energies $\sim$200 MeV/n. The helium
spectrum in the initial phase of the December 13 event is almost similar in
shape to the proton one, but flatter at energies below 200 MeV/n. In the event of December 13 the initial
spectra were rather hard and changed significantly at least till
1000 UT. Further evolution of the proton and helium spectra was
rather smooth even in spite of a Forbush effect in the middle of
December 14, until arrival at $\sim$2300 UT of the SEP accelerated in
the flare at 2107 UT on December 14. The intensity of $\sim$100 MeV protons
firstly increased till $\sim$0400 UT on December 13 and stayed at almost the same
level for at least 4 hours, while higher energy particles were rather
rapidly removed from the Earth's vicinity. After that, the particle
intensity decreased, but the spectrum shape was almost constant. Flux
intensity did not decrease to the solar quiet levels because of the
onset of December 14 event. During this event spectral form slightly flattened in the lower energy
part thus remaining quasi-exponential. The highest proton energy for
this event was around 500 MeV.

\begin{figure}
  \begin{center}
  \includegraphics[width=1.1\textwidth]{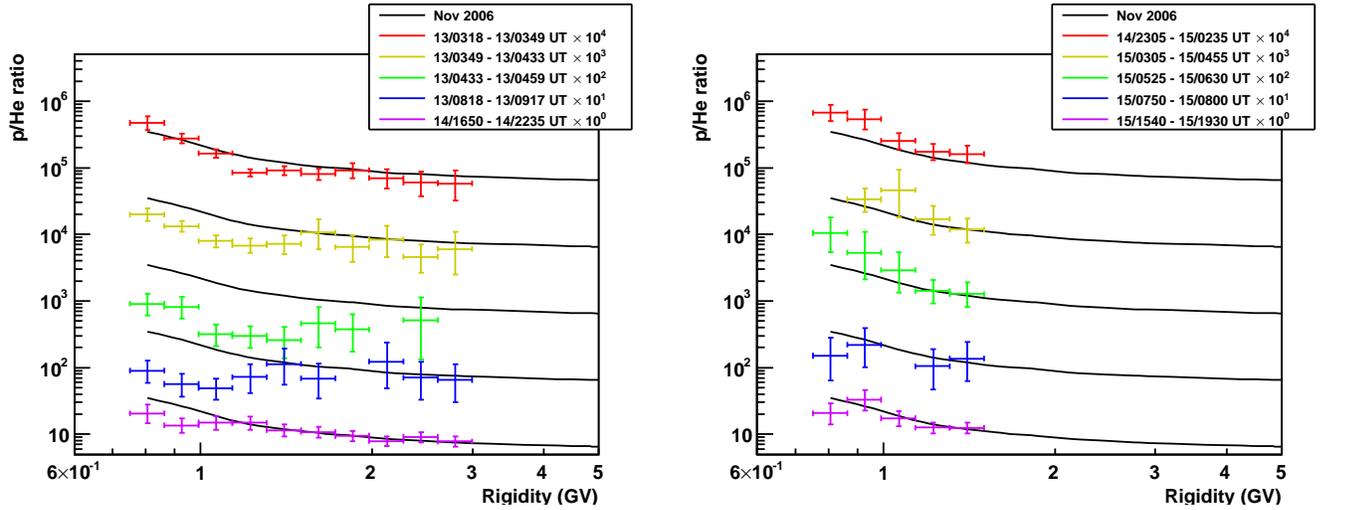}
  \end{center}
  \caption{The proton-to-helium ratio vs. rigidity at selected time intervals during the solar event of December 13 (left panel) and December 14 (right panel). The ratios are scaled to clarify the results. The GCR background is not subtracted and the ratio observed before the solar events is given as a black line} \label{fig_ratio}
\end{figure}

We have examined temporal behavior of proton-to-helium ratio in the both solar events. For this analysis the rigidity spectra were evaluated with rigidity intervals chosen in accordance with \cite{adriani_pamela_2011}. Figure \ref{fig_ratio} presents the proton-to-helium ratio obtained in selected time intervals. The galactic background was not subtracted and is shown as a black curves. It is seen in the left panel of  Figure \ref{fig_ratio} that at the beginning of the December 13 event the ratio was almost the same as for GCR (note that SEP intensity was highest at that time). The subsequent measurements showed the enhancement of He relative to protons up to the late December 14 when the high rigidity He virtually disappeared after subtraction of the galactic flux. Such a behavior may to a first approximation be interpreted as a result of faster removing of solar protons from the Earth’s orbit. Protons have higher velocity than He of the same rigidity therefore the diffusion coefficient of He at the same rigidity is smaller and protons leave faster.  Further analysis needs a particle propagation modeling. Right panel of Figure \ref{fig_ratio} clearly shows that no solar helium with rigidity above 1 GV was observed in the event of December 14.

No analysis directed to search of $^2H$ and $^3He$ has been performed yet.
No electrons or positrons of solar origin were found by the
\pam. However, it should be noted that the energy threshold of \pam\ is at $\sim$ 50 MeV, therefore rather high for solar electrons.

\subsection{Initial phase of the December 13, 2006 event}

The first arriving particles are objects of special interest because they provide more information about mechanism and conditions of acceleration. As it was mentioned, solar particles at the initial phase of the December 13 event demonstrated highly anisotropic angular distribution. In such conditions the asymptotic directions of particles arriving at PAMELA should be taken into account.

The geomagnetic field acts not only as rigidity analyzer but also as an angular analyzer of charged particles coming from space. Because of declining in the geomagnetic field a particle with certain rigidity above the geomagnetic cutoff $R_c$ can arrive at a given site only from a definite direction outside the magnetosphere (asymptotic direction). Therefore, at any point of its orbit PAMELA accepted particles with certain rigidity from only one asymptotic direction.

\begin{figure}
  \includegraphics[width=\textwidth]{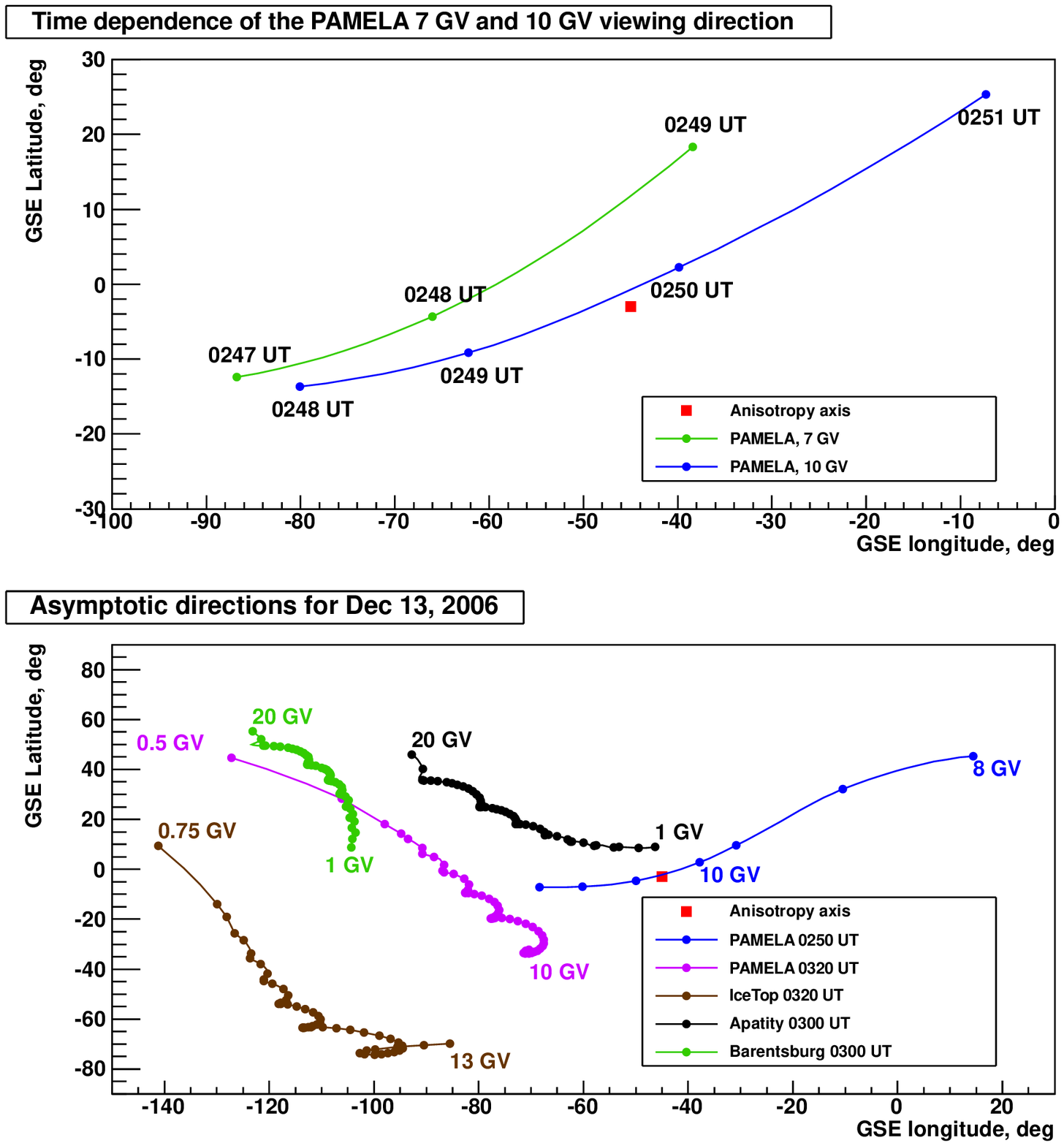}
  \caption{Upper panel: The asymptotic directions of 7 GV and 10 GV protons arriving at \pam\ in the initial phase of the SEP event at selected times of December 13, 2006 (given near the curves). Lower panel: Asymptotic directions of protons arriving at \pam\/, the Apatity and Barentsburg neutron monitors, the IceTop installation. The particle rigidity is marked near the corresponding direction. Red squares denote the direction of the SEP anisotropy axis.}
\label{fig_axis}
\end{figure}

At the moment of the first particle arrival \pam\ was at the
latitude of 40N moving to the equator. Therefore the low energy
particles could not reach the \pam\ detector. In addition, viewing
directions of \pam\ were changed rapidly while the spacecraft
moved along the orbit. This can be seen in the upper panel of
Figure \ref{fig_axis} where the asymptotic directions of 7 GV and
10 GV protons arriving at \pam\ are shown for the initial phase of
the SEP event. The asymptotic viewing directions of \pam\ has been
calculated with the program of \cite{gvozdevsky_b.b._private_2009}
using the back-tracing of solar particle trajectories and the
magnetospheric model of \cite{tsyganenko_model_2002}. It is in
reasonable agreement with \cite{plainaki2009AdSpR..43..518P}
obtained with the Tsyganenko 1989 model. The direction of the SEP
anisotropy axis, i.e. direction from which the bulk of SEP comes
is also plotted \citep{vashenyuk_characteristics_2008}. According
to Figure \ref{fig_axis}, \pam\ viewing directions at the early
phase of the SEP event would allow the solar protons in the range
of 7-10 GV to be recorded by \pam\ at 0248-0250 UT as these
particles arrived within a $\sim$30 degree cone around the SEP
flux axis. However, above 3 GeV the statistic is poor and we have
found no increase of proton flux there. Changes in the viewing
directions of \pam\ resulted in rapid receding of its favorable
direction of acceptance from the solar proton anisotropy axis. By
0256 UT (when a collimated particle bunch was observed by
\cite{timashkov_ground-level_2008}) the geomagnetic cutoff of
\pam\ was 12 GV and only protons with rigidity between 15 and 17
GV could come from directions close to the anisotropy axis.

The lower panel of Figure \ref{fig_axis} shows the asymptotic
directions of particles with different energies arriving at \pam\ in
the beginning of the event and at the most optimal time of observation
when the energy spectrum of particles with energies above $\sim$80
MeV/n was measured. For comparison, the asymptotic directions for the
Apatity neutron monitor and for the IceTop installation \citep{abbasi_solar_2008} are shown in
Figure \ref{fig_axis}.

\subsection{Comparison with other data}

\begin{figure}
  \begin{center}
  \subfloat[]{\includegraphics[width=0.55\textwidth,height=0.45\textwidth]{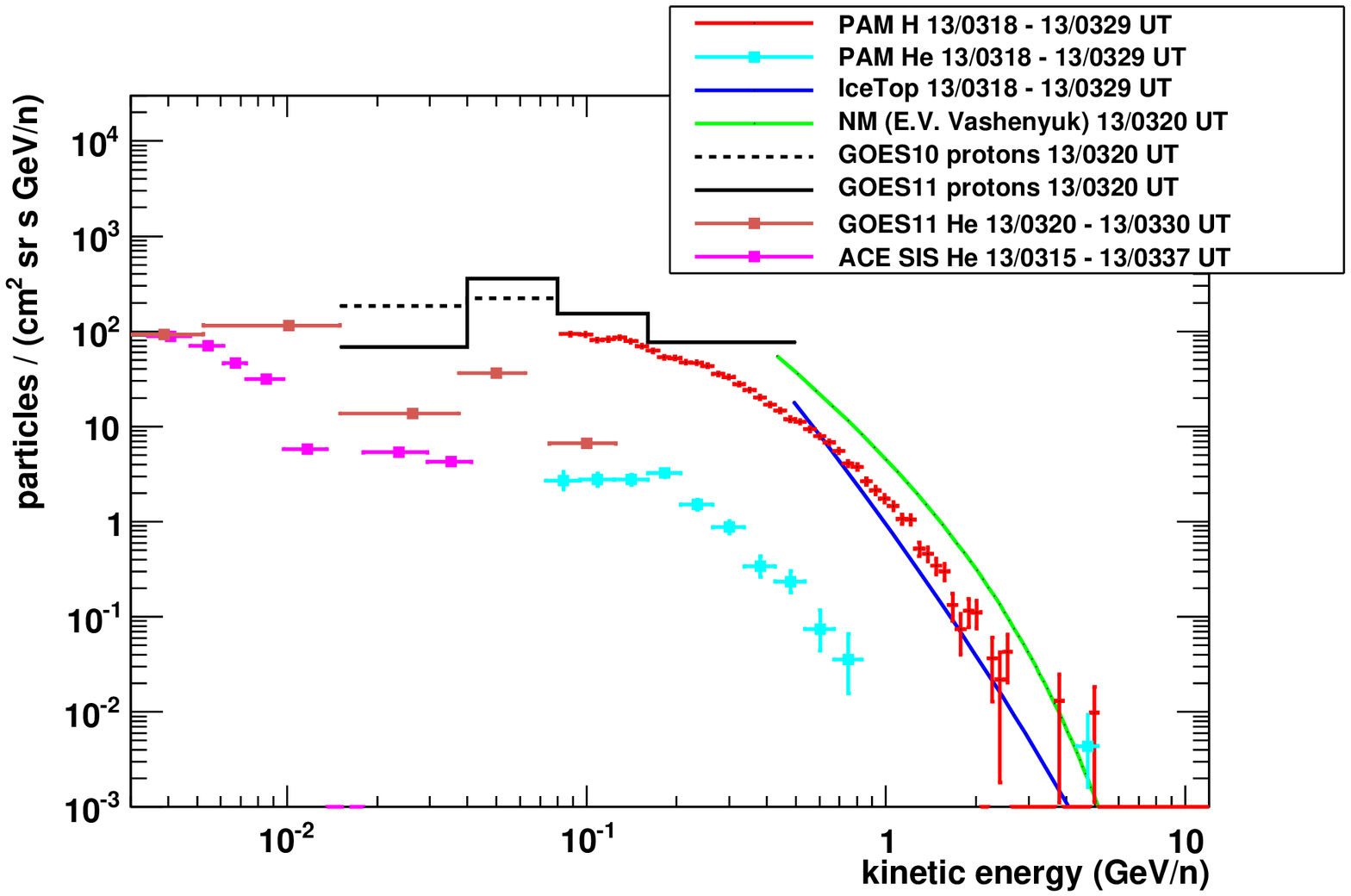}}
  \subfloat[]{\includegraphics[width=0.55\textwidth,height=0.45\textwidth]{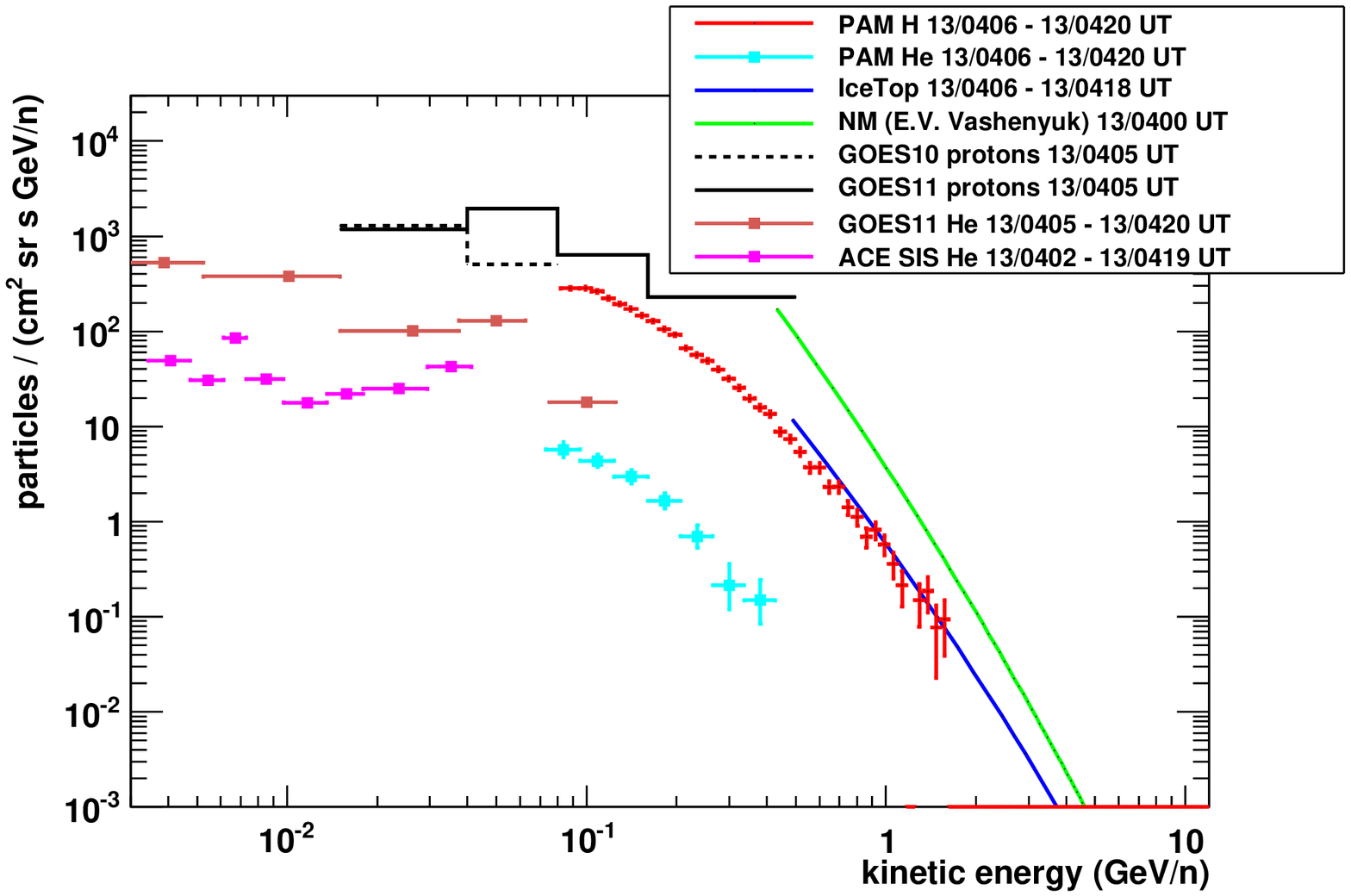}} \\
  \subfloat[]{\includegraphics[width=0.55\textwidth,height=0.45\textwidth]{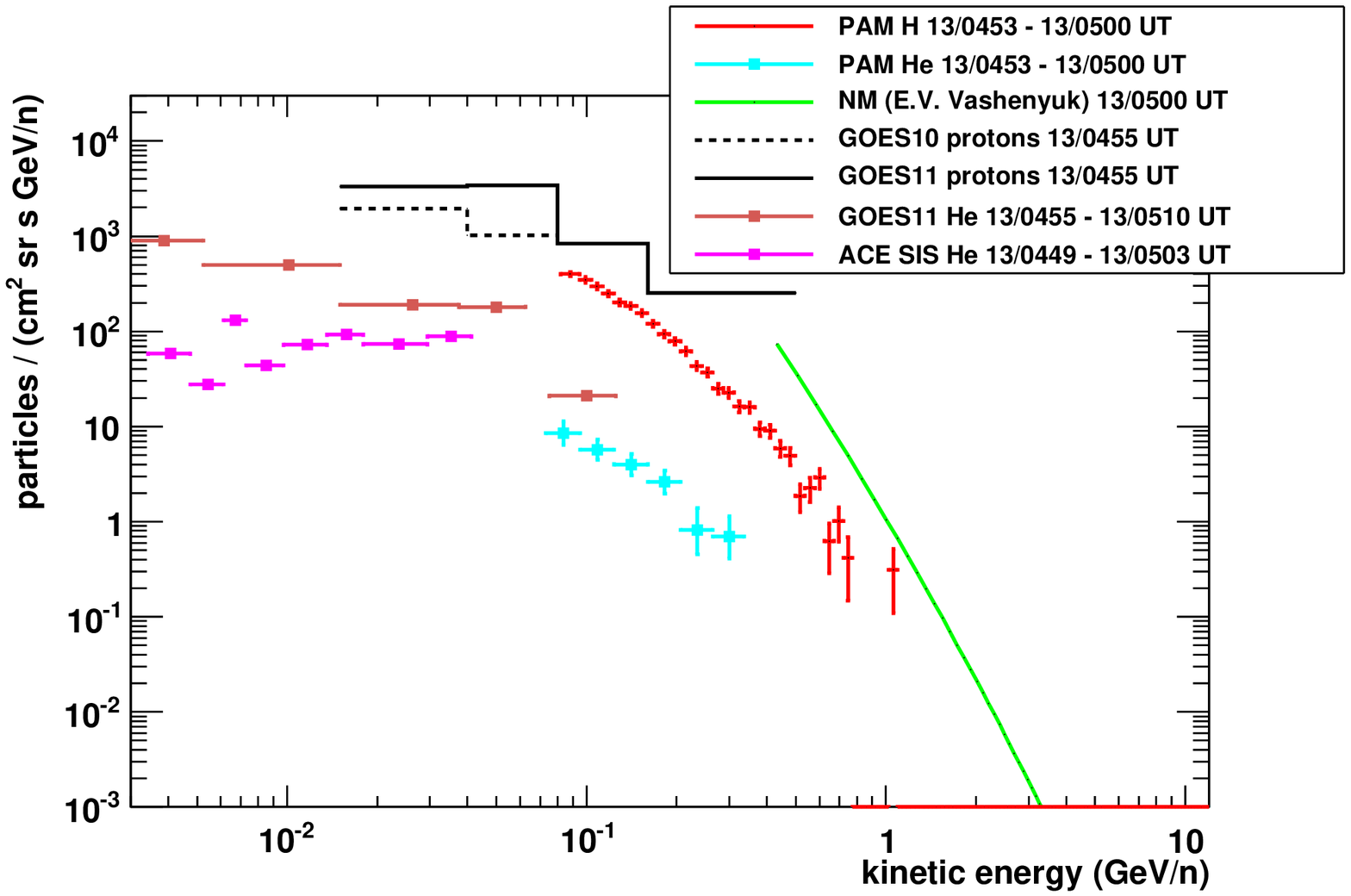}}
  \subfloat[]{\includegraphics[width=0.55\textwidth,height=0.45\textwidth]{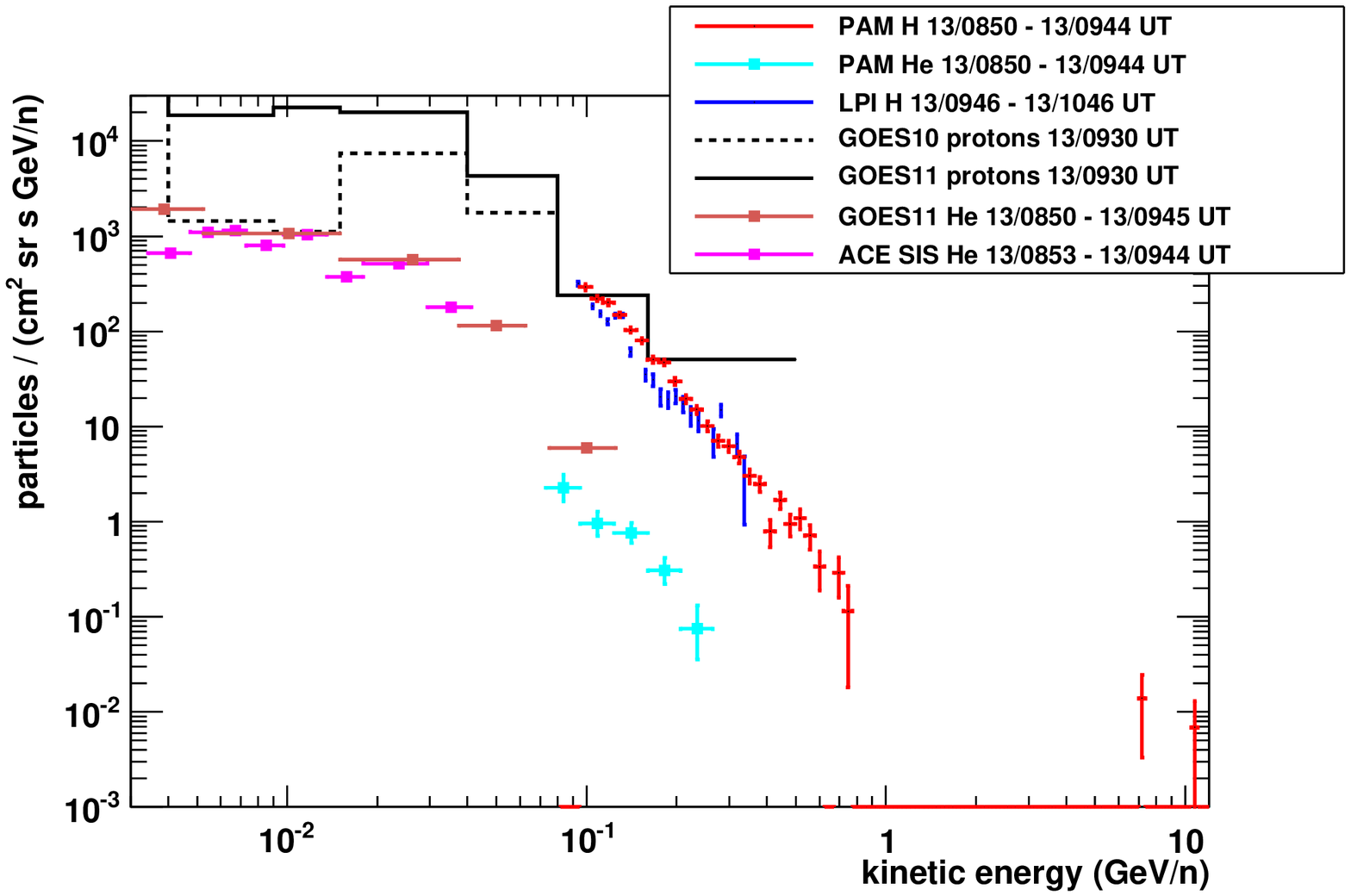}}
  \end{center}
  \caption{Energy spectra of solar particles measured by the \pam\ spectrometer (with galactic background subtracted) and by other experiments. In red there are the \pam\ protons, in cyan the \pam\ He, horizontal bars are the GOES and ACE SIS data. (a-c): green line is the spectrum derived from the neutron monitor network, blue line is the IceTop spectrum; (d): in blue there are the LPI balloon data.} \label{fig_comp}
\end{figure}

In Figure \ref{fig_comp} the differential spectra of \pam\, with the galactic background subtracted are compared with other experiments. Each panel corresponds to different time period.

In the low-energy range \pam\ particle fluxes can be
compared with other spacecraft measurements. The GOES \citep{GOES2011-1} fluxes are in moderate agreement with \pam\ ones, with exception for the most energetic GOES band (160-500 MeV) where GOES fluxes are several times higher. The same is true for GOES helium spectra. This could be due to contamination of lower energy particles in these GOES channels. On the other hand, the \pam\ helium fluxes are in a good agreement with measurements, performed by ACE/SIS \citep{ACE2011} in the adjacent energy interval.

At high energy, our data can be compared with the IceTop shower array \citep{abbasi_solar_2008} and the NM network\citep{vashenyuk_characteristics_2008}. The results of the IceTop shower array are plotted in Figure \ref{fig_comp}a, b specially taken at the time intervals of \pam\ observations \citep{kuwabara_t._private_2009}. The IceTop installation has detected
secondary particles, generated by the solar energetic particles,
i.e. the calculated yield functions have been used to reconstruct the
primary SEP fluxes. The IceTop results are depicted only in the energy
range over which this detector was actually seeing a significant number of
particles. The SEP fluxes obtained by \pam\ and
those obtained by IceTop is in relatively good agreement. However, at 0318-0329 UT (Figure \ref{fig_comp}a) the
power-law shape of the IceTop spectrum does not match the \pam\
spectrum which is more exponential-like. At 0406-0420 UT (Figure
\ref{fig_comp}b) the spectra of \pam\ and that of IceTop are virtually coincident in
the overlapping energy range.

The energy spectra from the NM network have been derived
from the early phase of the event (before 0300 UT on December 13) till
$\sim$ 0500 UT on December 13 \citep{balabin_yu.v.__2009}. As already mentioned, \pam\ missed the early phase of the event and took the first measurement at 0318-0329 UT,
during mildly anisotropic particle arrival (Figure \ref{fig_comp}a). A shape of the
\pam\ proton spectrum is close to that derived from the neutron
monitor network at energies between 700 MeV and 2 GeV. At lower energies
the \pam\ spectrum is harder; at $E\;>\; 2\,$ GeV the
uncertainty in the \pam\ spectrum is high because of small statistics.
The absolute intensities of protons from \pam\ are somewhat lower than those
derived from neutron monitors even at energies above 1 GeV. To some
extent, discrepancy between the \pam\ and neutron monitor spectra may
be caused by the flux anisotropy.

The GLE parameters for this event have also been calculated using
the NM-BANGLE model  \cite{NMBANGLE2009AdSpR..43..474P}. The
direct comparison of the particle fluxes is difficult since these
data in \cite{NMBANGLE2009AdSpR..43..474P} are given only for
integral energy spectra (see their Figure 3). An estimation shows
that the results of the NM-BANGLE model are very close to the
spectra obtained by \cite{vashenyuk_characteristics_2008} and
therefore differ from the PAMELA results in similar manner.
    Figure
\ref{fig_time_profile}, shows that during the first polar passage
of \pam\ (0318-0329 UT) the Apatity and Barentsburg neutron
monitors demonstrated different rate enhancements, i.e. the SEP
fluxes were still anisotropic. The neutron monitor spectrum refers
to the direction of the anisotropy axis where the proton intensity
is maximal. The \pam\ viewing directions for the protons below 3
GV (see lower panel of Figure \ref{fig_axis}) were not very close
to the anisotropy axis, therefore the \pam\ fluxes may be lower
than derived from neutron monitors. By the next \pam\ polar
passage (0406-0420 UT) discrepancy between the \pam\ and neutron
monitor spectra increased (Figure \ref{fig_comp}b). Since the flux
anisotropy becames lower, one should expect better consistency
between the \pam\ and neutron monitor data. On the contrary,
observations show discrepancy which persisted at around 0500 UT
(Figure \ref{fig_comp}c). Particle fluxes changed slowly at
0800-1000 UT, so in order to increase statistical significance,
the \pam\ results were averaged over 0850-0944 UT (Figure.
\ref{fig_comp}d). By this time the effect on the neutron monitor
network was already too small for determining the solar proton
spectrum. However, the \pam\ results were confirmed by an
independent observation of the LPI balloon-borne detector
\citep{stozhkov_y.i_fluxes_2007} which measured solar proton
energy spectrum in the stratosphere at 0946-1046 UT. Energy of
solar protons is determined in the LPI experiment from absorption
of protons in the air, so this is also a direct measurement of
solar protons although in the limited energy interval
\citep{bazilevskaya_solar_2010}. The good agreement between proton
data of the two instruments in the energy range 90 MeV - 400 MeV
is an additional check that our efficiency and systematics are
under control.

\section{The energy spectrum fitting}
The observed spectral shape may give some indications on
the acceleration mechanism \citep{casolino_two_2008}. The proton
energy spectrum measured in the early phase of the event, which is most
close to the time of primary acceleration (0318-0349 UT), was fitted by
the functions representative of various acceleration processes:

\begin{eqnarray}
\Phi_p = A\; e^{-E/E_0} \label{eq:expkine}\\
\Phi_p = A\; e^{-R/R_0} \label{eq:exprig}\\
\Phi_p = A\; R^{-\gamma-\delta(R-R_0) / R_0} \label{eq:plawmod}\\
\Phi_p = A\; R\;K_2\;(R / c\,\alpha\,T)^{1/2} \label{eq:bessel}
\end{eqnarray}
where $\Phi_p$ is the proton flux intensity, $E$ is kinetic energy, $R$ is magnetic rigidity, c is particle velocity, and $K_2$ is
a modified Bessel function of order 2, with $\alpha\;T$ as free
parameter ($\alpha$ representing an acceleration rate and $T$ the escape
time from the acceleration region). An exponential in kinetic energy (\ref{eq:expkine})
or rigidity (\ref{eq:exprig}) function is typical for simple models of DC acceleration
\citep{vashenyuk_relativistic_2006,vashenyuk_characteristics_2008,vashenyuk_relativistic_2008},
a power law is
indicative of shock acceleration
\citep{axford_acceleration_1981,krymskii_regular_1977,ellison_shock_1985},
and Bessel function is resulted from stochastic
acceleration \citep{mcguire_energy_1984}. The best fit parameters of
approximation are presented in Table~\ref{tabfits}.
\begin{table}[!h]
\centering
\begin{tabular}{|l|l|c|l|c|}

\hline
Function & A, $cm^{-2} sr^{-1} s^{-1} GV^{-1} (GeV^{-1})$ &
\multicolumn{2}{|c|}{Parameters} & $\chi^2 / ndf$\\
\hline
$A \cdot exp(-E/E_0)$ & $(68.2 \pm 0.9) $ & $E_0$, MeV & $(262 \pm 2)$ & 3.0\\
\hline
$A \cdot exp(-R/R_0)$ & $(179 \pm 3)$ & $R_0$, MV & $(366 \pm 2)$ & 7.6\\
\hline
\multirow{2}{*}{$A \cdot R^{-\gamma-\delta(R-R_0)/R_0}$, $R_0=1 GV$} & \multirow{2}{*}{$(13.1 \pm 0.1)$} & $\gamma$
& $(2.70 \pm 0.02)$ & \multirow{2}{*}{4.1}\\
\cline{3-4}
& & $\delta$ & $(1.99 \pm 0.06)$ & \\
\hline
$A \cdot R \cdot (K_2\,(R / c \cdot \alpha \cdot T))^{1/2}$ & $(106
\pm 2)$ & $\alpha \; T$ & $(1.71 \pm 0.01)$ & 6.4\\
\hline
\end{tabular}
\caption{The best fit parameters of the solar proton spectra in the energy interval 80 MeV -- 4 GeV in the early phase of the event (0318 - 0349 UT).}
\label{tabfits}
\end{table}

It should be mentioned that even if the resulting fits reproduce rather nicely
the spectral shape, $\chi^2$ values are high for all of them
 (see Table~\ref{tabfits}), showing that these analytical formulas do
not correctly describe the spectrum in the whole energy range. The best
fit is obtained for the exponential in kinetic energy function (\ref{eq:expkine}), however it
fails to describe the highest and lowest energy tails where the
observed spectrum is harder (in the $R\;>\; 2.5\, GV$ and $R\;<\;1.7$ GV range). It proves that no single mechanism can describe the SEP
energy spectrum in the wide energy interval. The previous successful
attempts of the SEP spectra fitting \citep[e.g.][]{cramp_october_1997,vashenyuk_characteristics_2008} were possible only for narrower energy range.

\section{Discussion and conclusion}

The \pam\ spectrometer was the first instrument which directly measured the
relativistic SEP in the near Earth space. It is important since all previous such measurements were fulfilled with the ground-based installation and the derived SEP fluxes depended on the instrument response functions. The spectra of solar protons in the energy range 80 MeV - 3 GeV and helium 75 MeV/n - 1 GeV/n were measured during the first polar \pam\ passage at 0318-0329
UT. There is a good agreement between the protons fluxes
measured by \pam\ and those obtained by the IceTop installation
\citep{abbasi_solar_2008,kuwabara_t._private_2009}. Keeping in mind accuracy of estimation
of absolute SEP intensities from the neutron monitor data
\citep{vashenyuk_relativistic_2006,plainaki_modeling_2007,bombardieri_improved_2008,NMBANGLE2009AdSpR..43..474P}, reasonable
agreement can be stated between the \pam\ and neutron monitor SEP
fluxes. However, the \pam\ spectra are always harder in the low-energy
interval probably indicating that neutron monitor yield functions are
underestimated below $\sim$700 MeV. During the second polar \pam\ passage the
difference between the SEP fluxes taken from \pam\ and neutron monitors
has become larger while agreement between \pam\ and IceTop remained
very good. The correctness of the \pam\ observations was also confirmed by
agreement with the direct SEP measurement on a balloon in the
atmosphere.

Evolution of intensity and spectral shape of relativistic solar protons
is often used to derive some information about SEP generation
\citep{bombardieri_improved_2008,vashenyuk_relativistic_2006,vashenyuk_characteristics_2008,vashenyuk_relativistic_2008,mccracken_two_2008, moraal_analysis_2008}. The indications were found that the first arriving relativistic particles would be accelerated in the flare region and have an exponential spectrum whereas the latter particles would be accelerated by a CME-driven shock and have a power-law spectrum. In this case an appropriate dynamics of the energy spectrum could be observed by \pam. Being at the low latitudes \pam\ missed the earliest anisotropic phase of the event of December 13. The results of the first \pam\ observation around 0320 UT on December 13 confirmed existence of a hard quasi-exponential spectrum expected from the magnetic reconnection in the flare region. However, in spite of changes in the SEP intensity a quasi-exponential spectrum persisted, actually till the advent of the newly generated protons of the December 14 event. It should be noted that in the energy interval relevant to the neutron monitor observations ($>$ 500 MeV) the spectrum could be fitted by a power law beginning from $\sim$1000 UT, however at lower energies the spectrum was flatter. We did not find any spectral form which would fit the observed spectra satisfactorily in the whole energy range and could prove a certain dominating mechanism of SEP acceleration. A quasi-exponential spectrum and fast temporal evolution of particle fluxes during several hours were present only in the event of December 13 when the relativistic protons were generated. In the event of December 14, without relativistic particles, a form of the solar proton energy spectra changed little and was almost power-law throughout the event. In addition, no Helium with energy above $\sim$100 MeV/n was observed in the December 14 event. This may be indicative of special conditions leading to acceleration of particles up to relativistic energy.

\section*{Acknowledgment}

We acknowledge support from The Italian Space Agency (ASI), Deutsches
Zentrum f\"ur Luft und Raumfahrt (DLR), The Swedish National Space Board, The Swedish Research Council, the Russian Space Agency (Roscosmos) and the Russian Foundation for Basic Research. We thank the IceTop team and the Polar Geophysical Institute cosmic ray group (Apatity) for providing the IceTop and the neutron
monitor spectra at the times of \pam\ measurements, the ACE SIS
\ instrument team and the ACE Science Center for providing the ACE
data. Our thanks are also to all people who make their results
accessible through the Internet.

\newpage

\bigskip

\end{document}